\documentclass[12pt]{article}

\usepackage{psfig}
\usepackage{overcite}
\begin{document}
\begin{center}
\Large{\bf Controlling the Short-Range Order and Packing Densities of Many-Particle Systems}

\vspace{0.2in}
\normalsize{}

S. Torquato$^{1,2}$ and F. H. Stillinger$^1$

 $^{1}$Department of Chemistry and $^{2}$Princeton Materials Institute \\
Princeton University, Princeton, NJ 08544 \\
\end{center}

\centerline{\bf Abstract}
\vspace{0.2in}

Questions surrounding the spatial disposition of particles in various condensed-matter systems continue to pose 
many theoretical challenges.  This paper explores the geometric availability of amorphous many-particle configurations 
that conform to a given pair correlation function $g(r)$.  Such a study is required to observe the basic constraints
of non-negativity for $g(r)$ as well as for its structure factor $S(k)$. The hard sphere case receives special attention, 
to help identify what qualitative features play significant roles in determining upper limits to maximum amorphous packing 
densities.  For that purpose, a five-parameter test family of $g$'s has been considered, which incorporates the known 
features of core exclusion, contact pairs, and damped oscillatory short-range order beyond contact.  Numerical optimization 
over this five-parameter set produces a maximum-packing value  for the fraction of covered volume, and about 5.8 for the mean
contact number, both of which are within the range of previous experimental and simulational  packing results.  However, the
corresponding maximum-density $g(r)$ and $S(k)$ display some unexpected characteristics. These include absence of any pairs 
at about 1.4 times the sphere collision diameter, and a surprisingly large magnitude for $S(k=0)$, the measure of  
macroscopic-distance-scale density variations. On the basis of these results, we conclude that restoration of more subtle
features to the test-function family of $g$'s (i.e. a split second peak, and a jump discontinuity at twice the collision 
diameter) will remove these unusual characteristics, while presumably increasing the maximum density slightly. A byproduct 
of our investigation is a  lower bound on the maximum density for random sphere packings in $d$ dimensions, which is 
sharper than a well-known lower bound for regular lattice packings for $d \ge 3$.

\section{Introduction}

Over a broad range of length
scales, many-particle systems  exhibit a rich variety
of structures with varying degrees of
long-range order, spanning from crystals, quasicrystals, and polycrystals
to amorphous solids and liquids. Consequently, it is natural
to focus attention on the statistical mechanics of the arrangement
of the particles. In the case of a macroscopic 
system containing a large number $N$ of particles, a full configurational description of that system usually is neither feasible, desirable, nor necessary.  For most practical purposes, 
it suffices to determine, or to describe, the distribution functions of 
low orders $n \ll N$.  Conventionally, this information is conveyed in the form of 
correlation functions.  For statistically homogeneous systems consisting
of identical spherical particles in a volume $V$, these correlation functions are defined so that 
$\rho^n g^{(n)}({\bf r}_1,{\bf r}_2,\dots,{\bf r}_n)$  is proportional to 
the probability density for simultaneously finding $n$ particles at 
locations ${\bf r}_1,{\bf r}_2,\dots,{\bf r}_n$ within the system,\cite{Hi56}
where $\rho=N/V$ is the number density. With this convention, each $g^{(n)}$ approaches 
unity when all particle positions become widely separated within V.

     The present study concerns the special circumstances for which 
the constituent particles are spherically symmetric and identical, and the system 
is statistically homogeneous and isotropic.  These conditions can be satisfied if the 
system contains a single fluid or amorphous solid phase.  The correlation 
function of primary interest is $g^{(2)}(r_{12})$, depending configurationally just on 
scalar pair distance $r_{12}$, and thus specifying how many pair distances of a 
given length occur statistically within the system.  The third-order 
function $g^{(3)}(r_{12},r_{13},r_{23})$ reveals how these pair separations are linked into triangles.  This additional information strictly speaking cannot be inferred from the 
knowledge of $g \equiv g^{(2)}$ alone, although the Kirkwood superposition 
approximation\cite{Hi56,Ki35} presumes to fill that knowledge gap.  The fourth-order 
function $g^{(4)}$ controls the assembly of triangles into tetrahedra, and is the 
lowest-order correlation function that is sensitive to {\it chirality} of the 
medium.

     On account of their probability interpretation, all of the $g^{(n)}$ must be 
non-negative; in particular, for all $r \ge 0$ we must have
\begin{equation}
g(r) \ge 0.
\label{g}
\end{equation}                                                              
In addition to this fundamental constraint, $g(r)$ is also subject to another 
basic inequality that arises from its connection to density 
fluctuations.  This concerns the behavior of the structure factor
defined  thus:\cite{Ha86}
\begin{eqnarray}
S(k)&=&1+\rho \int \exp(-i {\bf k} \cdot {\bf r}) [g(r)-1]d{\bf r} \nonumber\\
&=& 1+{4 \pi}\rho \int_{0}^{\infty} \frac{r\sin kr}{k}[g(r)-1] dr.
\label{S1}
\end{eqnarray} 
The second line assumes that we are treating three-dimensional systems.
The second fundamental constraint is the non-negativity of $S(k)$, i.e.,
\begin{equation}
S(k) \ge 0,
\label{S2}
\end{equation}
which must be obeyed for all real values of $k$.  It should be noted that 
(\ref{g}) and (\ref{S2}) are not at all restricted to states of thermal 
equilibrium, but are more general.  It is currently unknown if these 
necessary conditions (\ref{g}) and (\ref{S2})  are also sufficient to guarantee 
that any function  satisfying them is actually the pair correlation 
function for a realizable many-particle system; however, no 
counterexamples are currently known.

     Two recent studies\cite{St01,Sa02} have examined the theoretical possibility of 
controlling pair interactions in an isothermal many-particle system over 
a non-vanishing density range, starting at $\rho=0$, in such a way that $g(r)$ remains 
invariant over that range.  The cases examined have assigned forms to the 
invariant $g$ that were the zero-density limits appropriate for rigid rods, 
disks, and spheres,\cite{St01} and for the hard core plus square well pair 
potential.\cite{Sa02}  In each of these examples an upper terminal density $\rho^*$ could 
be identified such that over
\begin{equation}
0 \le \rho \le \rho^*
\end{equation}                                                                
the  $g$ invariance could indeed be maintained.  However, crossing $\rho^*$ would 
cause the $S(k)$ inequality (\ref{S2}) to be violated at some $k$.

     The principal objective of the present project is to apply the 
$g$-invariance technique to the still-challenging problem of random packings
of spheres. It has now been well established that the old concepts 
of  "random loose packing" and "random close packing" are ill-defined.\cite{To02}  
Instead, a non-trivial density range exists over which irregular packings 
of various types (locally, collectively, or strictly jammed\cite{To01}) can be 
formed, the preferred densities and local structures of which depend on the
preparation algorithm.  Our objective has been to study the logical 
connections between qualifying rigid-sphere $g$'s and the maximum 
corresponding densities, emerging as the terminal $\rho^*$'s.

     The following Section II introduces a parametric family of 
qualitatively reasonable, but functionally elementary, pair correlation 
functions for the amorphous-state sphere spatial arrangement problem.  
This family contains a set of five adjustable parameters, whose values 
must of course be consistent with the two basic inequalities (\ref{g}) and 
(\ref{S2}).  Section III describes a numerical search procedure over these 
parameters, and its results, the goal of which was to produce the largest 
$\rho^*$.  We believe it is significant that even with such a simple parametric 
family of pair correlation functions, $\rho^*$ can come close to that obtained in 
many experimental and simulational preparations of random sphere 
packings.  Section IV offers some interpretive remarks stimulated 
by the numerical results in Section III, and indicates the natural and 
useful directions for future investigation.  Finally, in an appendix, we derive
 a  lower bound on  the maximum density $\rho^*$ for random sphere
packings in $d$ dimensions and show that it is sharper than a well-known lower
bound for regular lattice packings for $d \ge 3$.

\section{Model Family}

Our interest in this paper is  to study models in which long-range
order is suppressed and short-range order is controlled.
Information that is available from previous determinations of local 
order in amorphous sphere packings provides useful guidance in choosing a 
model family of  functions for the present investigation.  In particular, 
we note that a survey of several experimental
and computer-simulation protocols,\cite{Fi70,Za83,Lu91} using distinct 
packing preparation procedures, appear to agree on the presence of  some 
qualitative features. (Figure \ref{lub} shows the pair correlation function
for a dense random packing of spheres as generated by us employing
the Lubachevsky-Stillinger ``compression'' protocol.\cite{Lu91}) Using the sphere-pair distance of closest approach 
as the natural length unit, these $g(r)$ attributes are the following:

\begin{itemize}
\item[(i)]  Obviously, $g(r)$ must vanish for all $0 \le r \le 1$.
\item[(ii)]  On account of the jamming, virtually all spheres (a few 
"rattler" spheres can be present as exceptions) are rigidly in contact 
with neighbors.  The number of such contacts must average at least 4 to 
meet the definition of ``local" jamming.\cite{To01}
\item[(iii)]  For $r \ge 1$,  $g(r)$ displays finite-amplitude oscillations about unity, that 
decay to zero with increasing $r$.  The length scale of these oscillations 
is roughly comparable to the sphere diameter.
\item[(iv)]  A pair of distinctive $g(r)$ peaks appear at distances approximately 
equal to $\sqrt{3}$ and $\sqrt{4}$.  These are often termed a ``split second peak", and appear 
in modified form for amorphous deposits of soft-sphere and attracting 
particles.\cite{LJ}
\item[(v)]  As $r$ increases through $r=2$, $g(r)$ experiences a discontinuous drop in 
magnitude.
\end{itemize}

     It is not the objective of the present work to try to include all of 
these features slavishly.  Instead, we have chosen as a first step to 
represent only attributes (i), (ii), and (iii) by a simple parametric 
function family, and to see how close the largest corresponding density would 
come to the approximate experimental range\cite{Ha66,Po97}
of ``random'' packing densities\cite{footnote1}: 
\begin{equation}
1.18 \le \rho \le 1.26
\end{equation}
The equivalent approximate range of covering or packing fractions  $\phi\equiv \pi \rho/6$ is:
\begin{equation}
0.62 \le \phi \le 0.66
\end{equation}                                                        
Computer-simulation determinations of random packing fractions\cite{To02} lie in the wider
range
\begin{equation}
0.60 \le \phi \le 0.68,
\end{equation}
illustrating the fact that the packing densities
are protocol-dependent\cite{To00,To02}.
After the fact, this approach should determine how important the 
remaining attributes (iv) and (v) are for the random sphere packing problem.
     
Consequently, we have elected to write  $g(r)$ as a linear combination of 
three portions, corresponding respectively to (i), (ii), and (iii) above:
\begin{equation}
g(r)=g_I(r)+g_{II}(r)+g_{III}(r).
\label{sum}
\end{equation}
The first involves just the unit step function U:                
\begin{equation}
g_I(r)=U(r-1),
\end{equation}
while the second represents the sphere contacts:
\begin{equation}
g_{II}(r)=\frac{Z}{4\pi\rho}\delta(r-1).
\end{equation}
Here, $Z$ is the first of our adjustable parameters, equal to the mean 
number of contacts (coordination number) experienced by each sphere.  The third portion 
contains four more adjustable dimensionless parameters ($A$, $B$, $C$, $D$):
\begin{equation}
g_{III}(r)=\frac{A}{r}\exp(-Br) \sin(C r+D)U(r-1),
\end{equation}
and is intended to represent approximately the damped oscillation
beyond contact.                            

     The Fourier transforms required to evaluate the structure factor,
\begin{equation}
S(k)=1+\rho[G_I(k)+G_{II}(k)+G_{III}(k)]
\label{S}
\end{equation}               
are expressible entirely in terms of elementary functions:
\begin{eqnarray}
G_I(k)&=&\int \exp(-i {\bf k} \cdot {\bf r}) [g_I(r)-1]d{\bf r} \nonumber\\
&=& \frac{4 \pi}{k^3}  [k\cos k-\sin k]
\label{G1}
\end{eqnarray}
\begin{eqnarray}
G_{II}(k)&=&\int \exp(-i {\bf k} \cdot {\bf r}) [g_{II}(r)-1]d{\bf r} \nonumber\\
&=& \frac{Z}{\rho} \frac{\sin k}{k}
\label{G2}
\end{eqnarray}              
\begin{eqnarray}
G_{III}(k)&=&\int \exp(-i {\bf k} \cdot {\bf r}) [g_{III}(r)-1]d{\bf r} \nonumber\\
&=& \frac{2\pi\exp(-B)}{k}\Bigg[\frac{B\cos(k-C-D)-(k-C)\sin(k-C-D)}{B^2+(k-C)^2} \nonumber \\
&&- \frac{B\cos(k+C+D)-(k+C)\sin(k+C+D)}{B^2+(k+C)^2}\Bigg].
\end{eqnarray}

     The five adjustable parameters $Z$, $A$, $B$, $C$, and $D$  are subject to 
some obvious constraints.  Clearly, we must demand that 
\begin{equation}
Z \ge 0.
\label{Z}
\end{equation}               
In addition, exponential increase with $r$ is not permissible, so
\begin{equation}
B \ge 0.
\label{B}
\end{equation}
Furthermore, $D$ only needs to span a single period of the trigonometrical 
factor in which it occurs:
\begin{equation}
0 \le D \le 2\pi.
\label{D}
\end{equation}

The remaining parameter pair $A$, $C$ is not entirely free, of course, but 
must be consistent with both inequalities (\ref{g}) and (\ref{S2}).  Our central 
objective is to search over the five-dimensional domain defined by (\ref{g}), 
(\ref{S2}), and (\ref{Z})-(\ref{D}), for the maximum terminal density $\rho^*(Z,A,B,C,D)$
or terminal covering volume fraction $\phi^*(Z,A,B,C,D)$ that they 
permit.  Under the working assumption that this maximum is attained at 
the boundary of the five-dimensional domain, it becomes important to know 
which constraint or constraints are at issue there, and why.

\section{Numerical Search Procedure and Results}

The problem of finding the maximal packing fraction $\phi^*(Z,A,B,C,D)$ can
be posed as an optimization problem. 
Specifically, this can be posed as a two-level
``min-max'' problem: one wants to maximize $\phi(Z,A,B,C,D)$ over the parameters
$Z,A,B,C,$ and $D$ with the restrictions (\ref{Z})-(\ref{D})
such that the minimum of  $S(k;Z,A,B,C,D)$ 
in the variable
$k$ and the minimum of $g(r;Z,A,B,C,D)$ in the variable $r$ are
both non-negative, i.e.,
\begin{equation}
\max_{Z,A,B,C,D} \phi
\end{equation}
such that 
\begin{equation}
\min_{k} S(k;Z,A,B,C,D) \ge 0, \qquad \min_{r} g(r;Z,A,B,C,D) \ge 0.
\end{equation}
The interval-arithmetic paradigm\cite{Ke95} is a global optimization
methodology that in principle should
enable one to obtain exact narrow interval bounds on the maximal 
packing fraction in a computationally efficient manner. 
We attempted this calculation using the GlobSol Fortran 90 global
optimization library \cite{web} but could not
obtain an exact interval solution. This program is best suited
for finding conventional extrema
of simple differentiable functions with simple constraints.
Our problem is considerably more complex and so we instead used GlobSol
to find the minimum
\begin{displaymath}
\min_{k} S(k;Z,A,B,C,D)
\end{displaymath}
and then employed a brute-force grid search over $Z,A,B,C,$ and $D$
subject to the aforementioned conditions in order to maximize $\phi$.

The search procedure is implemented for six different cases summarized
in Table 1. In the first case (Case I), no restrictions on
the functions, other than the ones described above, are imposed.
In the remaining cases, we impose additional restrictions.
In particular, it is known that dense random packings
are typically spatially uniform, implying that $S(k=0)\approx 0$.
Therefore, in some instances, we  carry out the search 
subject to the condition that   $S(k=0)=0$. 
\bigskip

\noindent{\it Case I}
\smallskip

Not surprisingly, the least restrictive case yields the largest value of
the maximal
packing fraction: $\phi^*=0.627$. The corresponding values
of the parameters are listed in Table 2. Figure \ref{case1} shows the
structure factor and pair correlation
function. These functions reveal structural
features that are not characteristic of typical dense
random packings obtained either experimentally or
computationally. For example, the structure factor
at $k=0$ is unusually high, implying significant
density fluctuations in the infinite-wavelength limit.
Moreover, at three different finite wavelengths 
($k \approx 3.1$, $k \approx 6.3$, and $k \approx 10$)
the structure factor is essentially zero
or nearly zero, implying vanishingly small density
fluctuations at these wavelengths.
Atypically, the pair correlation function $g(r)$ exhibits appreciable oscillations
before attaining its long-range value
at about $r=8$. This feature could be the result
of polycrystallinity, but the fact that $g(r)$ vanishes
at $r \approx 1.4$ (another anomalous characteristic) evidently eliminates
the possibility that such putative crystallites
are face-centered cubic arrangements.
Interestingly, the coordination number $Z \approx 5.8$ is
approximately equal to the value observed in experiments and
simulations of typical dense
random packings.\cite{To02,Fi70,Za83} Theoretical arguments
have been put forth  predicting that $Z=6$ for
random packings of identical frictionless spheres in three
dimensions.\cite{Al98,Ed99,Wi99} 
\bigskip

\noindent{\it Case II}
\smallskip

In the second case, we conduct the search under the
condition that $S(k=0)=0$. For this condition to
hold, $Z$ must be given by
\begin{equation}
Z=8\phi-1- \frac{24\phi}{B^2+C^2}\Bigg[
\left(B+\frac{B^2-C^2}{B^2+C^2}\right)\sin(C+D)
+\left(C+\frac{2BC}{B^2+C^2}\right)\cos(C+D)\Bigg].
\end{equation}
This condition also implies that the term in $S(k)$ of
order $k^2$ is zero, and therefore the
first non-zero term is of order $k^4$.
Both the maximal packing fraction and coordination
number drop from the first case to the values $\phi^*=0.46$
and $Z=2.3964$ (see Tables 1 and 2).
Figure \ref{case2} shows the structure factor and pair correlation
function. Note the atypical curvature of the function $g(r)$ near the
contact value.

\bigskip
\noindent{\it Case III}
\smallskip

In the third case, we suppress the damped-oscillating component
[$g_{III}(r)=G_{III}(k)=0$]. Here we find that
$\phi^*=0.41 $ and $Z=3.1504 $.

\bigskip
\noindent{\it Case IV}
\smallskip

In the fourth case, we suppress the damped-oscillating component
[$g_{III}(r)=G_{III}(k)=0$] and we also impose the
condition $S(k=0)=0$. This problem can be solved exactly.
Here we find
\begin{equation}
\phi^*=\frac{5}{16}, \qquad Z= \frac{3}{2}.
\end{equation}
In the Appendix, we obtain the $d$-dimensional generalization
of this result and show how it leads to a lower bound
on $\phi^*$ for the general case in which the third
component is not suppressed [$g_{III}(r)\ne 0$, $G_{III}(k)\ne 0$].

\bigskip
\noindent{\it Case V}
\smallskip

In the fifth case, we suppress the delta-function component ($Z=0$).
Here we obtain $\phi^*=0.375$ (see Tables 1 and 2).

\bigskip
\noindent{\it Case VI}
\smallskip

In the sixth case, we suppress the delta-function component ($Z=0$) and we
also impose the
condition $S(k=0)=0$. We find that $\phi^*=0.3535$ (see Tables
1 and 2).

\section{Concluding Remarks}

For the family of pair correlation functions specified
by relation (8), the optimal packing fraction is characterized by
unusual structural features such as substantially large
density fluctuations in the infinite-wavelength limit, vanishing
or nearly vanishing fluctuations at several finite wavelength values, 
and an interparticle radial distance ($r \approx 1.4$)
at which particle centers 
are prohibited. Nonetheless, the maximal packing fraction
and coordination number ($\phi^*=0.627$ and $Z=5.8$) 
are consistent with values  for
dense random packings generated experimentally and computationally.
Clearly, however, the properties (iv) and (v)
(``split second peak" and the  discontinuous drop at $r=2$) that
are characteristic of typical random packings  are absent
in the optimal solution. In future work, one may want to consider
other families of functions that are not as smooth as (8) away from $r=1$, e.g.,
piecewise continuous functions. Such extensions would quantify
the significance of properties (iv) and (v) in raising $\phi^*$
above the optimal value of $0.627$, while presumably driving
$S(k=0)$ downward toward zero.

Assuming that the optimal packing can actually be realized,
there remain many open questions.
Do the spheres form a contacting percolating network?  Given the
high density that is achieved, we suspect that the answer is
in the affirmative. If so, what is the geometry of the contact
set? Are rattlers present in the optimal solution?
Is the packing locally, collectively, or strictly jammed?\cite{To01}
The answers to all of these questions can greatly be facilitated
if we could determine whether there are packings that achieve the optimal solution.
This can be accomplished using stochastic reconstruction
techniques that enable one to obtain realizations of
sphere packings that have a targeted pair correlation function
or structure factor\cite{Ri97}.
We will attempt such a reconstruction in
a future study.
Another interesting extension of the present work
is to generalize the family of pair correlation functions
to the case of spheres of different sizes
and to determine the  maximal packing fraction.

\noindent{\bf Acknowledgements}

The authors are grateful to
Aleksandar Donev for his help with the
interval-arithmetic calculations.
S. T. was supported by
the Petroleum Research Fund as administered
by the American Chemical Society.
\vspace{0.4in}

\noindent{\Large\bf Appendix: $d$-dimensional Generalization of Case IV}
\bigskip

In this Appendix, we obtain an exact expression for the optimal
value of $\phi^*$ for the $d$-dimensional generalization of Case IV.
A consequence of this result is a  lower bound on  $\phi^*$ for random sphere
packings in $d$ dimensions, which we compare to a well-known lower
bound for regular lattice packings.

Consider the evaluation of the structure factor $S(k)$ of relation (\ref{S})
in $d$ dimensions but without the short-ranged (damped-oscillating) 
contribution, i.e.,
\begin{equation}
S(k)=1+\rho[G_I(k)+G_{II}(k)]
\label{S-d}
\end{equation}
where $G_I(k)$ and $G_{II}(k)$ are the $d$-dimensional Fourier transforms~\cite{Sn95}
\begin{equation}
    G_I(k) =
\left(2\pi\right)^{\frac{d}{2}}\int_{0}^{\infty}r^{d-1}\left\{g_I(r)-1\right\}
 \frac{J_{\left(d/2\right)-1}\!\left(kr\right)}{\left(kr\right)^{\left(d/2\right)-1}}
\,\,dr,
  \end{equation}
\begin{equation}
    G_{II}(k) =
\left(2\pi\right)^{\frac{d}{2}}\int_{0}^{\infty}r^{d-1}\left\{g_{II}(r)-1\right\}
\frac{J_{\left(d/2\right)-1}\!\left(kr\right)}{\left(kr\right)^{\left(d/2\right)-1}}\,\,dr,
  \end{equation} 
\begin{equation}
g_I(r)=U(r-1),
\end{equation}
\begin{equation}
g_{II}(r)=\frac{Z}{s_1(1)\rho}\delta(r-1),
\end{equation}
and
\begin{equation}
s_1(r)  =  \frac{2\pi^{d/2}r^{d-1}}{\Gamma(d/2)}.
\label{area-sph}
\end{equation}
The quantity $s_1(r)$
is the surface area of a  $d$-dimensional sphere of radius $r$.

It immediately follows that
\begin{equation}
\rho G_I(k)= -\frac{2^{3d/2}\Gamma(1+d/2) \phi}{(k)^{d/2}} J_{d/2}(k),
\label{g1}
\end{equation}
where
\begin{equation}
\phi= \rho \frac{\pi^{d/2}}{\Gamma(1+d/2)} (1/2)^d
\end{equation}
is the $d$-dimensional packing fraction. If this were the only contribution
to the structure factor,
then the non-negativity condition $S(k) \ge 0$ implies
\begin{equation}
\phi^* =\frac{1}{2^d},
\end{equation}
which agrees with the result given in Ref. 4. It easily
follows that
\begin{equation}
\rho G_{II}(k)= \frac{2^{d/2}\Gamma(1+d/2)Z }{d(k)^{d/2-1}} J_{d/2-1}(k),
\label{g2}
\end{equation}
Substitution of (\ref{g1}) and (\ref{g2}) into (\ref{S-d}) yields
\begin{equation}
S(k)=1+(Z-2^d\phi)+\left[\frac{2^{d-2}\phi}{1+d/2}-\frac{Z}{2d}\right] k^2+
{\cal O}(k^4)
\end{equation}
The last term changes sign if $Z$ increases past $2^d \phi d/(d+2)$.
At this crossover point,
\begin{equation}
S(k)=1-\frac{2^{d+1}}{d+2} \phi+ {\cal O}(k^2)
\end{equation}
Since the minimum occurs at $k=0$, then we have the exact results
\begin{equation}
\phi^*=\frac{d+2}{2^{d+1}}, \qquad Z=\frac{d}{2}.
\end{equation}
Thus, we have the lower bound  
\begin{equation}
\phi^* \ge \frac{d+2}{2^{d+1}}
\label{lower}
\end{equation}
for random sphere packings
in $d$ dimensions because the addition of the short-range contribution
(which we have neglected) would result in a generally larger value
of $\phi^*$. In obtaining lower bound (\ref{lower}), we have assumed
there are no further sufficiency conditions
beyond (\ref{g}) and (\ref{S2}).

In very high dimensions ($d \sim 1000$), the densest known
packings are non-regular lattices.\cite{Co93} 
Thus, it is of interest to compare the lower bound (\ref{lower}) to 
the Minkowski-Hlawka theorem,\cite{Co93} which 
gives a lower bound on the maximum
packing fraction for $d$-dimensional regular lattices of identical spheres:
\begin{equation}
\phi_{max} \ge \frac{\zeta(d)}{2^{d-1}},
\label{bound}
\end{equation}
where $\zeta(d)$ is the Riemann zeta function. The lower bound 
(\ref{lower}) is larger than the lower bound (\ref{bound}) for $d \ge 3$.
Indeed, the difference between these bounds grows with increasing $d$.

\newpage
\begin{table}[h]
\centering
\caption{Terminal packing fractions $\phi^*$ for six different cases
in which the step-function contribution $g_I$ to $g$ in (\ref{sum})
is always included.\label{packing}} 
\begin{tabular}{|c|c|c|c|c|}
\multicolumn{5}{c}{~} \\\hline 
Case I & $g_{II}\ne 0$  & $g_{III}\ne 0 $ & $S(k) > 0$ &  $\phi^*=0.627$ \\ \hline 
Case II &$g_{II}\ne 0$  & $g_{III}\ne 0 $ & $S(k) = 0$ &  $\phi^*=0.46$ \\ \hline 
Case III &$g_{II}\ne 0$  & $g_{III}= 0 $ & $S(k) > 0$ &  $\phi^*=0.41$ \\ \hline 
Case IV &$g_{II}\ne 0$  & $g_{III}= 0 $ & $S(k) = 0$ &  $\phi^*=0.3125$ \\ \hline 
Case V &$g_{II}= 0$  & $g_{III}\ne 0 $ & $S(k) > 0$ &  $\phi^*=0.375$ \\ \hline 
Case VI & $g_{II}= 0$  & $g_{III}\ne 0 $ & $S(k) = 0$ &  $\phi^*=0.3535$ \\ \hline 
\end{tabular}
\end{table}

\hspace{0.1in}
\vspace{1.0in}

\begin{table}[h]
\centering
\caption{Values of the parameters for the cases in Table 1.\label{packing2}} 
\begin{tabular}{|c|c|c|c|c|c|}
\multicolumn{6}{c}{~} \\\hline 
Case I & $A=2.733$  & $B=0.510 $ & $C=7.471 $ &  $D=0.627 $ & $Z=5.80$\\ \hline 
Case II & $A=1.15$  & $B=0.510 $ & $C=5.90 $ &  $D=1.66 $ & $Z=2.3964$\\ \hline
Case III & $A=0$  & $B=0 $ & $C=0 $ &  $D=0 $ & $Z=3.1504$\\ \hline
Case IV & $A=0$  & $B=0 $ & $C=0 $ &  $D=0 $ & $Z=1.5$\\ \hline
Case V & $A=4.8$
 & $B=1.2 $ & $C=5.90 $ &  $D=0.90 $ & $Z=0$\\ \hline
Case VI & $A=3.9$  & $B=0.9 $ & $C=5.70 $ &  $D=0.90 $ & $Z=0$\\ \hline
\end{tabular}
\end{table}

\newpage

\begin{center}
{\bf Figure Captions}
\end{center}

\begin{figure}[h]
\caption{The pair correlation
function  $g(r)$ vs. $r$ as obtained by
averaging over  ten configurations at a packing fraction $\phi = 0.64$.
The ``binned'' peak value of $g(1)$ (not shown) is approximately 24.}
\label{lub}
\end{figure}

\begin{figure}[h]
\caption{Structure factor (a) and pair correlation
function (b) for Case I. Note the appearance of a vertical line at 
contact in (b) indicating a delta-function contribution
there.}
\label{case1}
\end{figure}

\begin{figure}[h]
\caption{Structure factor (a) and pair correlation
function (b) for Case II.  Note the appearance of a vertical line at 
contact in (b) indicating a delta-function contribution
there.}
\label{case2}
\end{figure}

\newpage

\begin{figure}
%\centerline{\psfig{file=g2.eps,width=6in}}
\centerline{\psfig{file=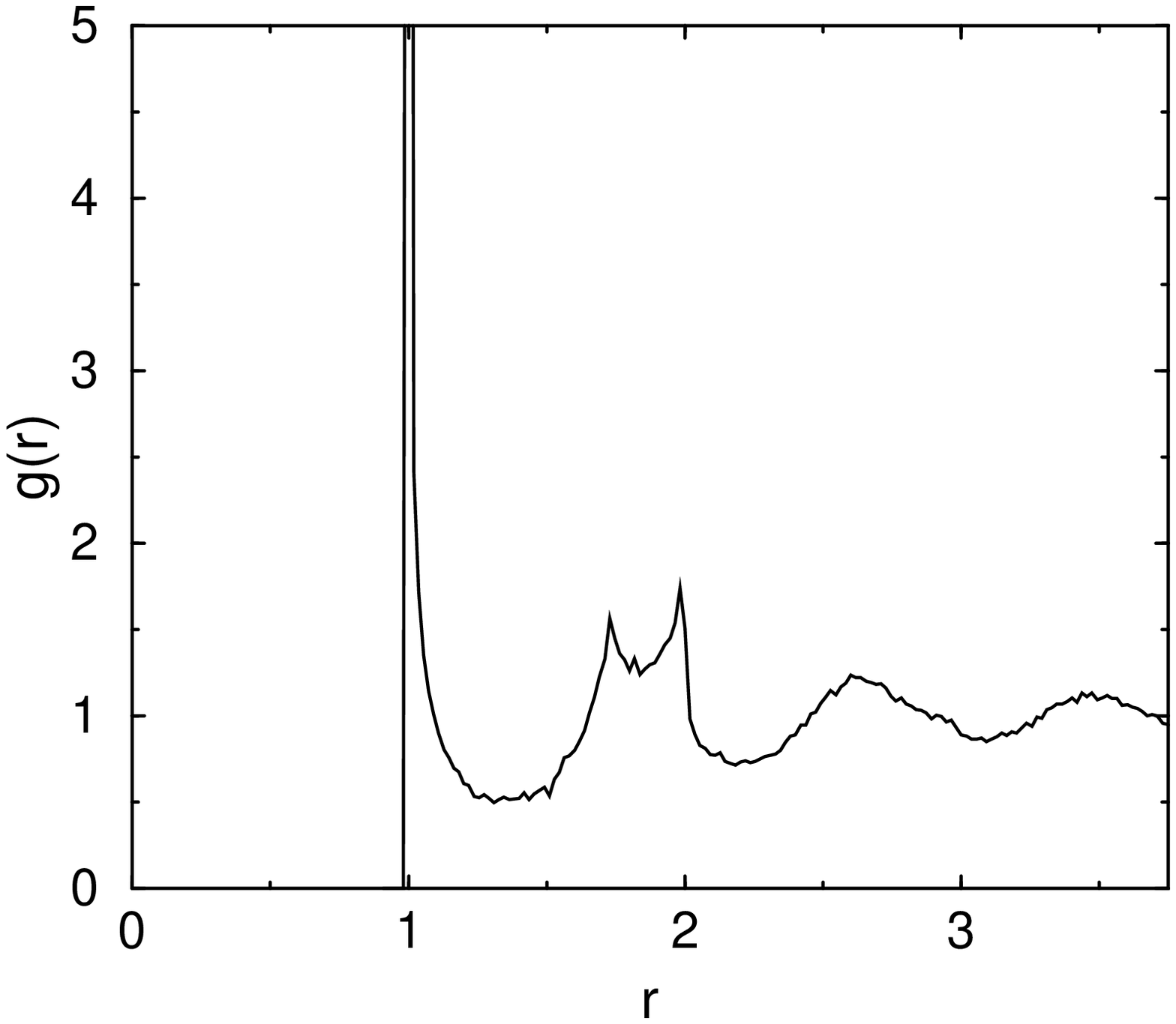,width=6in}}
\end{figure}
\vspace{0.4in}

\centerline{\large\bf Figure 1}

\newpage

\begin{figure}
%\centerline{\psfig{file=S-high.eps,width=6in}}
\centerline{\psfig{file=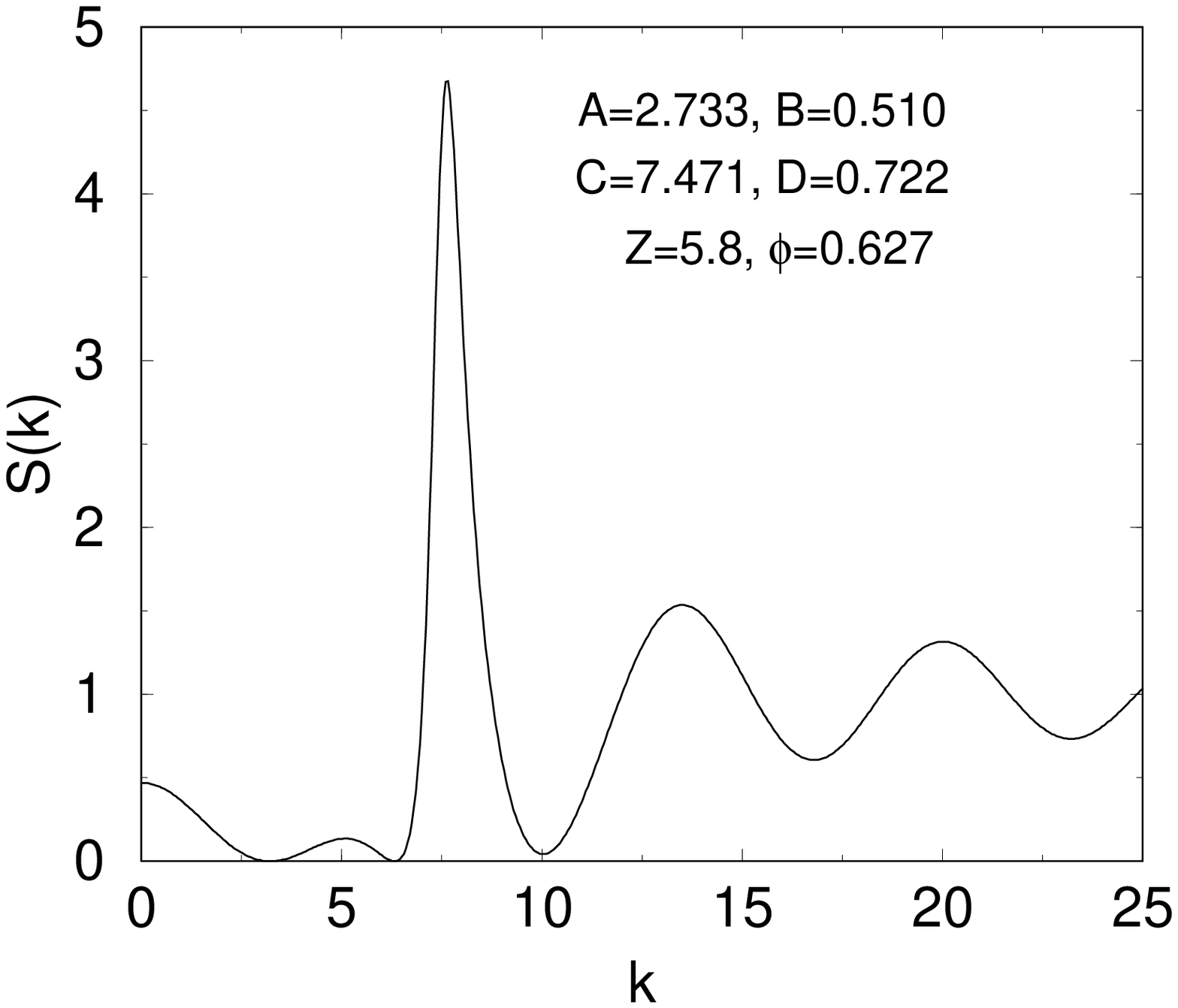,width=6in}}
\end{figure}
\vspace{0.4in}

\centerline{\large\bf Figure 2a}

\newpage

\begin{figure}
%\centerline{\psfig{file=g-high.eps,width=6in}}
\centerline{\psfig{file=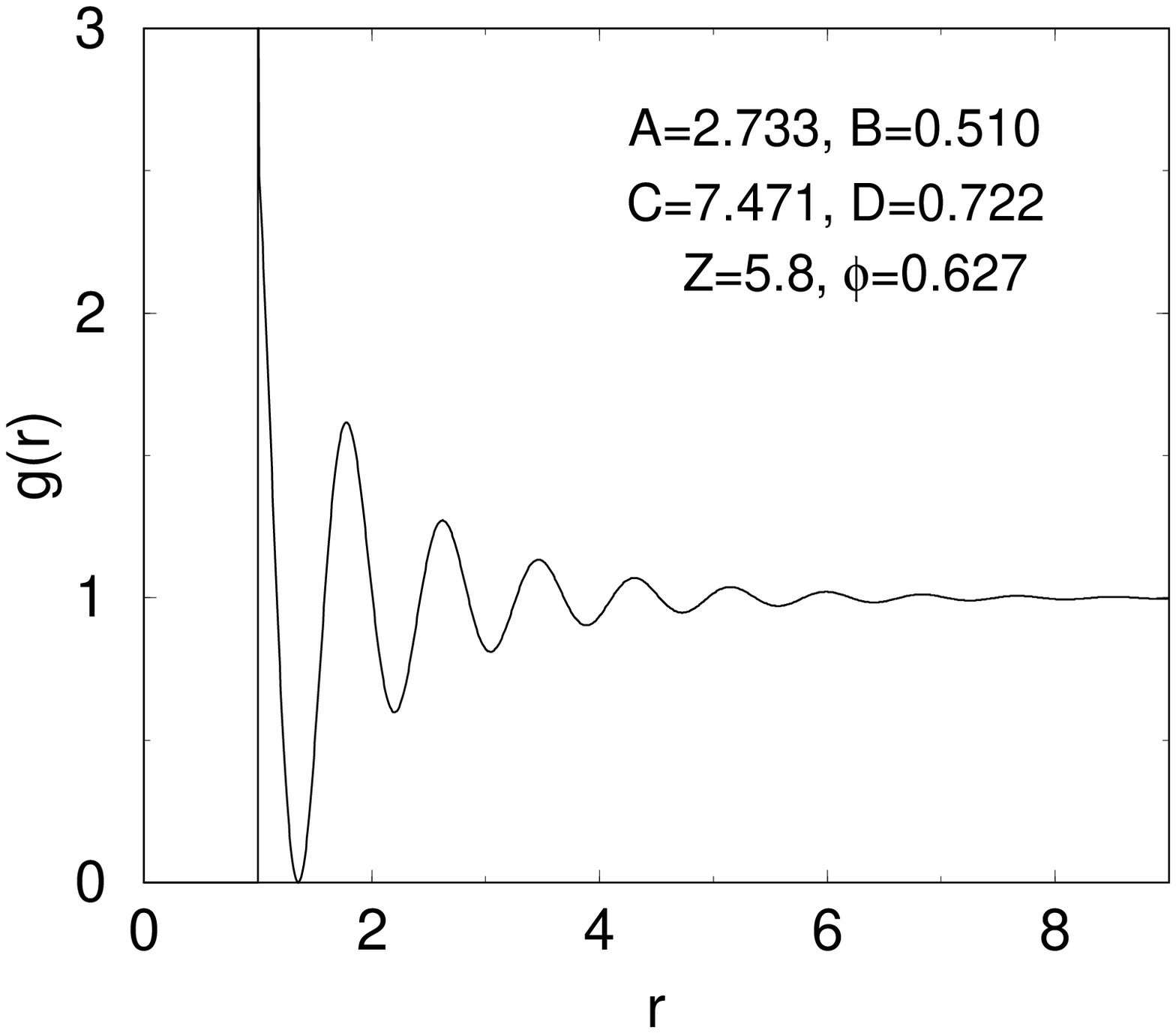,width=6in}}
\end{figure}
\vspace{0.4in}

\centerline{\large\bf Figure 2b}

\newpage

\begin{figure}
%\centerline{\psfig{file=S-zero.eps,width=6in}}
\centerline{\psfig{file=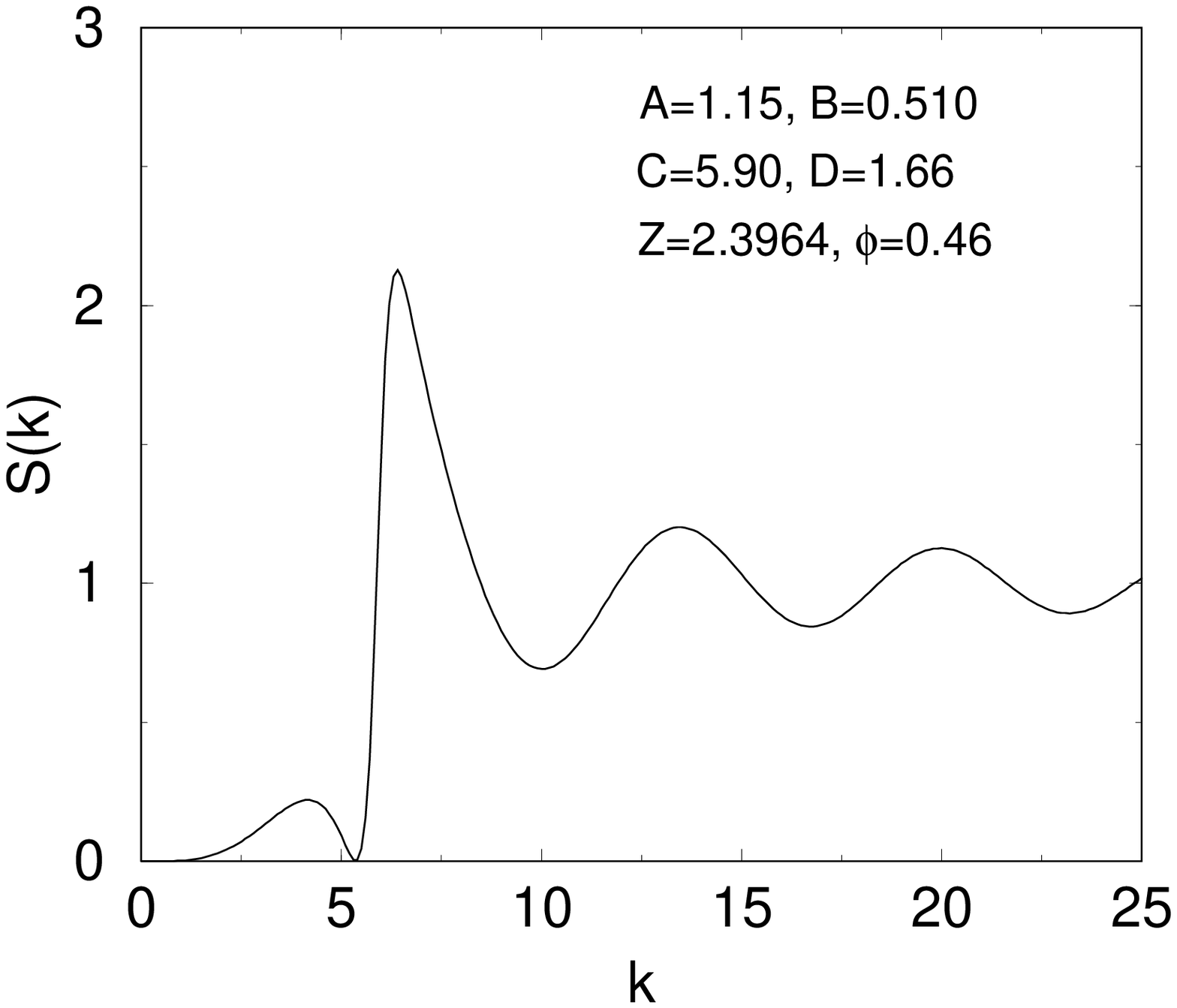,width=6in}}
\end{figure}
\vspace{0.4in}

\centerline{\large\bf Figure 3a}

\newpage

\begin{figure}
%\centerline{\psfig{file=g-zero.eps,width=6in}}
\centerline{\psfig{file=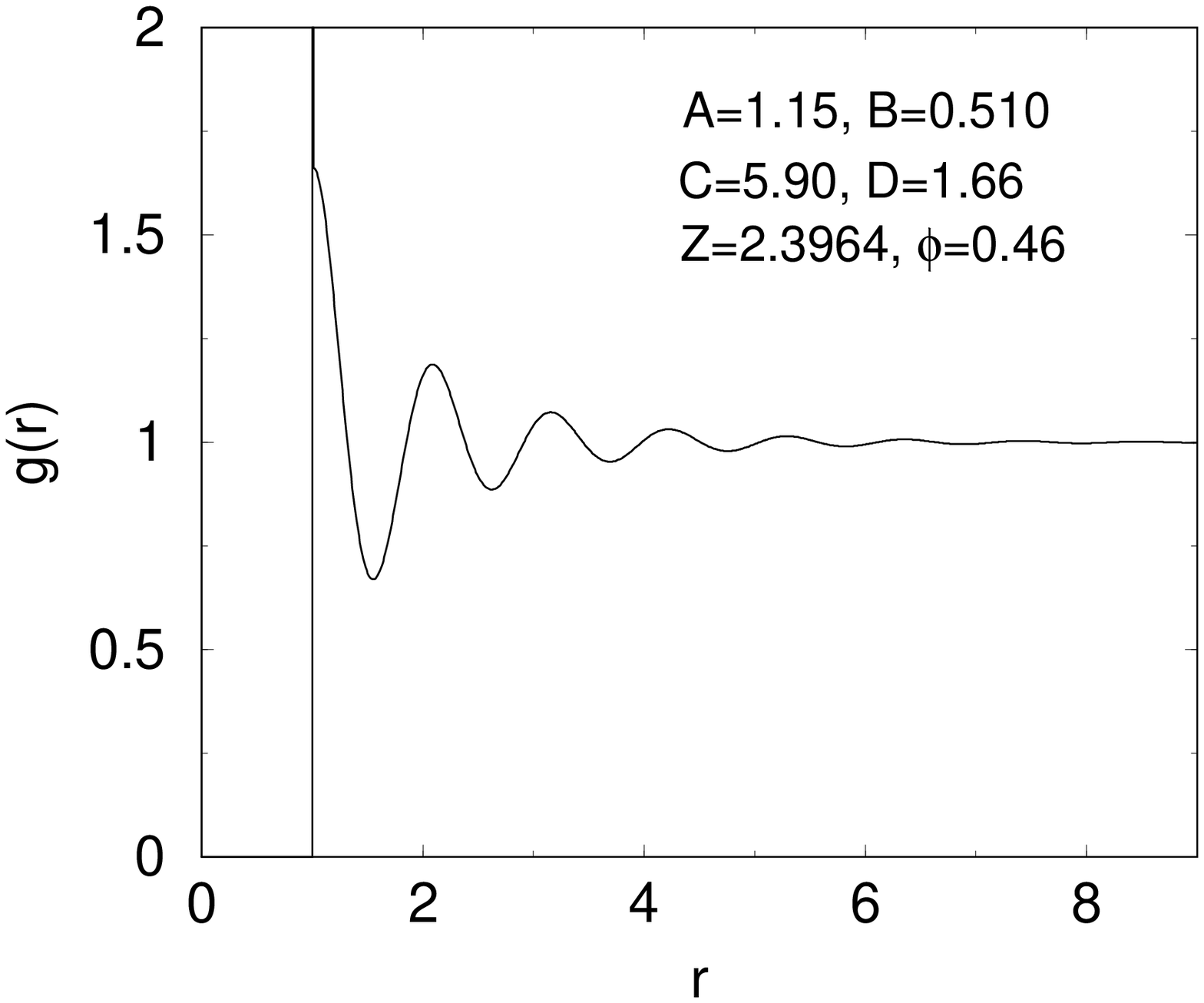,width=6in}}
\end{figure}
\vspace{0.4in}

\centerline{\large\bf Figure 3b}


\begin{thebibliography}{20}
\bibitem{footnote1} Note that we are careful not to refer to any of these states
as random close packing (RCP) because the RCP concept is not well-defined
mathematically, as discussed in Refs. 6 and 7.
\end{thebibliography}
\end{document}